# Polarization squeezing and continuous-variable polarization entanglement

Natalia Korolkova,[1] Gerd Leuchs,[1] Rodney Loudon,[1,2] Timothy C. Ralph[3] and Christine Silberhorn[1]

[1]Lehrstuhl für Optik, Physikalisches Institut, Friedrich–Alexander–Universität, Staudtstr. 7/B2, D–91058 Erlangen, Germany
[2]Department of Electronic Systems Engineering, Essex University, Colchester CO4 3SQ, UK
[3]Department of Physics, University of Queensland, St Lucia, QLD 4072, Australia

The Stokes-parameter operators and the associated Poincaré sphere, which describe the quantum-optical polarization properties of light, are defined and their basic properties are reviewed. The general features of the Stokes operators are illustrated by evaluation of their means and variances for a range of simple polarization states. Some of the examples show polarization squeezing, in which the variances of one or more Stokes parameters are smaller than the coherent-state value. The main object of the paper is the application of these concepts to bright squeezed light. It is shown that a light beam formed by interference of two orthogonally-polarized quadrature-squeezed beams exhibits squeezing in some of the Stokes parameters. Passage of such a primary polarization-squeezed beam through suitable optical components generates a pair of polarization-entangled light beams with the nature of a two-mode squeezed state. The use of pairs of primary polarization-squeezed light beams leads to substantially increased entanglement and to the generation of EPR-entangled light beams. The important advantage of these nonclassical polarization states for quantum communication is the possibility of experimentally determining all of the relevant conjugate variables of both squeezed and entangled fields using only linear optical elements followed by direct detection.

# I. INTRODUCTION

The *classical* Stokes parameters [1] provide a convenient description of the polarization properties of light and the complete range of the classical states of polarization is readily visualized by the use of the Poincaré sphere [2]. The *quantum* Stokes parameters [3–5] provide operator representations of the polarization that also apply to nonclassical light. The operators satisfy quantum-mechanical commutation relations and the variances of the Stokes parameters are accordingly restricted by uncertainty relations. The quantum states of polarization are conveniently visualized by an appropriate quantum version of the Poincaré sphere.

We consider a beam of light whose plane wavefronts are perpendicular to the $z$ axis and whose polarization lies in the $xy$ plane. The polarization state with quantum-mechanical coherent-state excitations of both the $x$ and $y$ polarization components has characteristic uncertainties that separate the classical and nonclassical regimes. Light is said to be *polarization squeezed* if the variance of one or more of the Stokes parameters is smaller than the corresponding value for coherent light. Methods to generate polarization-squeezed light using propagation through an anistropic Kerr medium have been proposed [5–9] (see [10] for a review). Frequency-tunable polarization-squeezed light, produced by combining the squeezed-vacuum output of an optical parametric oscillator with an orthogonally-polarized strong coherent beam on a polarizing beam splitter, has been applied to quantum state transfer from a light field to an atomic ensemble, thus generating spin squeezing of the atoms in an excited state [11].

The purpose of the present paper is to extend the general theory of the quantum Stokes parameters and Poincaré sphere and to propose straightforward experiments for generating and detecting polarization squeezing and entanglement. The basic properties of the Stokes parameters are outlined in Sec. II and these are illustrated in Sec. III by consideration of some simple idealized examples of polarization states. The quantum Stokes parameters of primary light beams whose two polarization components are formed from the more practical bright amplitude-squeezed light are evaluated in Sec. IV. A particular experimental scheme is outlined. Linear optical schemes for measuring the means and variances of all three parameters of the primary light beam by direct detection alone are outlined. The methods resemble those used for determination of the classical Stokes parameters, except that simultaneous measurements of different polarization components are needed for observation of the quantum effects. The measurement procedure produces a two-beam squeezed-state entanglement. Section V considers the EPR entanglement of the Stokes parameters that can be obtained by combination of two primary light beams with similar polarization characteristics.

The applications of nonclassical polarization states in quantum information communication and cryptography are discussed in Sec. VI.

## II. QUANTUM STOKES PARAMETERS AND POINCARE SPHERE

The Hermitian Stokes operators are defined as quantum versions of their classical counterparts [1,2]. Thus in the notation of [10],

$$\hat{S}_0 = \hat{a}_x^\dagger \hat{a}_x + \hat{a}_y^\dagger \hat{a}_y = \hat{n}_x + \hat{n}_y = \hat{n}, \tag{2.1}$$

$$\hat{S}_1 = \hat{a}_x^\dagger \hat{a}_x - \hat{a}_y^\dagger \hat{a}_y = \hat{n}_x - \hat{n}_y, \tag{2.2}$$

$$\hat{S}_2 = \hat{a}_x^\dagger \hat{a}_y + \hat{a}_y^\dagger \hat{a}_x, \tag{2.3}$$

$$\hat{S}_3 = i\left(\hat{a}_y^\dagger \hat{a}_x - \hat{a}_x^\dagger \hat{a}_y\right), \tag{2.4}$$

where the $x$ and $y$ subscripts label the creation, destruction and number operators of quantum harmonic oscillators associated with the $x$ and $y$ photon polarization modes, and $\hat{n}$ is the total photon-number operator. The creation and destruction operators have the usual commutation relations,

$$\left[\hat{a}_j, \hat{a}_k^\dagger\right] = \delta_{jk} \quad j,k = x, y. \tag{2.5}$$

The Stokes operator $\hat{S}_0$ commutes with all the others,

$$\left[\hat{S}_0, \hat{S}_i\right] = 0 \quad i = 1, 2, 3 \tag{2.6}$$

but the operators $\hat{S}_1$, $\hat{S}_2$ and $\hat{S}_3$ satisfy the commutation relations of the SU(2) Lie algebra, for example

$$\left[\hat{S}_2, \hat{S}_3\right] = 2i\hat{S}_1. \tag{2.7}$$

Apart from the factor of 2 and the absence of Planck's constant, this is identical to the commutation relation for components of the angular momentum operator. Simultaneous exact measurements of the quantities represented by these Stokes operators are thus impossible in general and their means and variances are restricted by the uncertainty relations

$$V_2 V_3 \geq \left|\left\langle \hat{S}_1 \right\rangle\right|^2, \quad V_3 V_1 \geq \left|\left\langle \hat{S}_2 \right\rangle\right|^2 \quad \text{and} \quad V_1 V_2 \geq \left|\left\langle \hat{S}_3 \right\rangle\right|^2. \tag{2.8}$$

Here $V_j$ is a convenient shorthand notation for the variance $\langle \hat{S}_j^2 \rangle - \langle \hat{S}_j \rangle^2$ of the quantum Stokes parameter $\hat{S}_j$.

It is readily shown [3] that

$$\hat{S}_1^2 + \hat{S}_2^2 + \hat{S}_3^2 = \hat{S}_0^2 + 2\hat{S}_0 \tag{2.9}$$

and this is taken to define the quantum Poincaré sphere. The mean value of the sphere radius is given by the square root of the expectation value of either side of (2.9) and it generally has a nonzero variance.

The relations (2.1) to (2.4) are equivalent to the well-known Schwinger representation of angular momentum operators in terms of a pair of quantum harmonic oscillators [12-15]. The quantum numbers $l$ and $m$ of the angular momentum state are related to the quantum numbers $n_x$ and $n_y$ of the harmonic oscillators by

$$l = \tfrac{1}{2}(n_x + n_y) \quad \text{and} \quad m = \tfrac{1}{2}(n_x - n_y). \tag{2.10}$$

A pure state of the polarized light field is denoted $|\psi; x, y\rangle$ and a density-operator description is needed for statistical mixture states. Some simple examples of pure state are treated in the following section to show the main characteristic features of the quantum Stokes parameters and Poincaré sphere.

## III. SIMPLE POLARIZATION STATES

*Number states*

Consider first the state of linearly-polarized light that has $n$ photons with $x$ polarization and no photons with $y$ polarization,

$$|\psi; x, y\rangle = |n\rangle_x |0\rangle_y. \tag{3.1}$$

The state is an eigenstate of the first two Stokes parameters, with

$$\hat{S}_0 |\psi; x, y\rangle = \hat{S}_1 |\psi; x, y\rangle = n |\psi; x, y\rangle. \tag{3.2}$$

Thus

$$\langle \hat{S}_0 \rangle = \langle \hat{S}_1 \rangle = n \quad \text{and} \quad V_0 = V_1 = 0. \tag{3.3}$$

The other two parameters have zero expectation values,

$$\langle\hat{S}_2\rangle = \langle\hat{S}_3\rangle = 0. \tag{3.4}$$

and the expectation values of their squares are

$$\langle\hat{S}_2^2\rangle = \langle\hat{S}_3^2\rangle = n = V_2 = V_3. \tag{3.5}$$

The state is, however, an eigenstate of the sum of these squared Stokes parameters, with

$$(\hat{S}_2^2 + \hat{S}_3^2)|\psi;x,y\rangle = 2n|\psi;x,y\rangle. \tag{3.6}$$

The uncertainty relations in (2.8) are all satisfied as equalities for the number state.

Figure 1 shows two sections of the Poincaré sphere for the $x$ polarized number state. The radius of the sphere has a well-defined value in view of the relations (3.2) and (3.6). The tip of the Stokes vector $(S_1, S_2, S_3)$ lies on a circle perpendicular to the $S_1$ axis at coordinate $S_1 = n$. The figure is identical, apart from some factors of 2, to that for an angular momentum vector with a well-defined $S_1$ component.

The number state is an eigenstate of the squared Stokes parameters in (3.5) for the special case of $n = 1$, when

$$\hat{S}_2^2|\psi;x,y\rangle = \hat{S}_3^2|\psi;x,y\rangle = |\psi;x,y\rangle \quad \text{for} \quad n = 1. \tag{3.7}$$

The corresponding angular momentum state in this case, given by (2.10), has $l,m$ quantum numbers $\tfrac{1}{2},\tfrac{1}{2}$ and the Pauli spin matrices accordingly provide a representation for the Stokes operators $\hat{S}_1$, $\hat{S}_2$ and $\hat{S}_3$.

*Coherent states*

Just as the photon-number polarization state is an analogue of the angular momentum state with well-defined magnitude and $S_1$ component, the coherent polarization state is an analogue of the coherent angular momentum, spin or atomic state [13,16,17]. With both photon polarization modes excited in independent coherent states, we denote the combined product state by

$$|\psi;x,y\rangle = |\alpha_x\rangle_x |\alpha_y\rangle_y = \hat{D}_x(\alpha_x)\hat{D}_y(\alpha_y)|0\rangle_x|0\rangle_y, \tag{3.8}$$

where $\hat{D}_j(\alpha_j)$ $j = x,y$, is the usual coherent-state displacement operator. The state is a simultaneous eigenstate of the mode destruction operators $\hat{a}_x$ and $\hat{a}_y$ with eigenvalues $\alpha_x$ and $\alpha_y$ respectively. The expectation values of the quantum Stokes parameters are then obtained by replacing the creation and destruction operators in (2.1) to (2.4) by $\alpha_j^*$ and $\alpha_j$ as appropriate [18], for example

$$\langle \hat{S}_0 \rangle = |\alpha_x|^2 + |\alpha_y|^2 = \langle \hat{n}_x \rangle + \langle \hat{n}_y \rangle = \langle \hat{n} \rangle. \tag{3.9}$$

The coherent-state complex amplitudes $\alpha_x$ and $\alpha_y$ correspond to the amplitudes used in the definitions of the classical Stokes parameters. The various possible states of polarization are specified by exactly the same values of $\alpha_x$ and $\alpha_y$ as in the classical theory [2].

In contrast to the classical theory, however, the radius of the Poincaré sphere is ill-defined because of uncertainties in the values of all the Stokes parameters. Their variances are all equal for the coherent states [5,10],

$$V_j = \langle \hat{n}_x \rangle + \langle \hat{n}_y \rangle = \langle \hat{n} \rangle \quad j = 0,1,2,3; \tag{3.10}$$

they bear the same relation to the mean value of $\hat{S}_0$ in (3.9) as do the photon-number variance and mean for the coherent state [19]. It is readily verified that the three uncertainty relations in (2.8) are satisfied and they take the forms of equalities for appropriate values of the coherent-state amplitudes and phases. The Poincaré sphere relation (2.9) is verified in the form

$$\langle \hat{S}_1^2 + \hat{S}_2^2 + \hat{S}_3^2 \rangle = \langle \hat{n}^2 + 2\hat{n} \rangle = \langle \hat{n} \rangle^2 + 3\langle \hat{n} \rangle \tag{3.11}$$

and the variance in the squared radius of the sphere is nonzero. The quantum Poincaré sphere for the coherent polarization state is therefore fuzzy, in contrast to that for the number state. Figure 2 shows two sections of the Poincaré sphere, which are drawn for the mean radius obtained from the square root of (3.11). Here $\alpha_y$ is set equal to zero for ease of comparison with Fig. 1. It is seen that, because of the equal variances (3.10) of the three Stokes parameters, the uncertainty is now represented by the shaded sphere of radius $\sqrt{3\langle n \rangle}$ centred on the mean value $(\langle n \rangle, 0, 0)$ of the Stokes vector.

The Poincaré sphere has the well-defined surface of its classical counterpart only in the limit of very large mean photon numbers, $\langle n \rangle \gg 1$, corresponding to bright coherent light, where the uncertainties in the Stokes parameters are negligible in comparison to the mean amplitude of the Stokes vector. The radius of the uncertainty sphere in Fig. 2 then shrinks relative to that of the Poincaré sphere and the tip of the Stokes vector approaches the surface of the Poincaré sphere.

Light is said to be *polarization squeezed*, according to the definition in Sec. I, when the variance in one or more of the Stokes parameters is smaller than the coherent-state value,

$$V_j < \langle \hat{n} \rangle \quad j = 1, 2, 3. \tag{3.12}$$

The Stokes parameter with $j = 0$ is excluded, as the condition (3.12) in this case is the same as that for photon-number squeezing. The photon-number state is polarization squeezed in the $S_1$ Stokes parameter according to (3.3), although this too is equivalent to photon-number squeezing.

*Entangled single-photon state*

Consider a number state defined as in (3.1) but now for a single photon excited with polarization in a direction $x'$ that bisects the $x$ and $y$ axes, and no photons excited with polarization in the orthogonal $y'$ direction. The state can be written

$$|\psi; x, y\rangle = |1\rangle_{x'} |0\rangle_{y'} = \hat{a}_{x'}^\dagger |0\rangle = 2^{-1/2} \left( \hat{a}_x^\dagger + \hat{a}_y^\dagger \right) |0\rangle = 2^{-1/2} \left( |1\rangle_x |0\rangle_y + |0\rangle_x |1\rangle_y \right), \tag{3.13}$$

where $|0\rangle$ is the two-dimensional vacuum state. The resulting state in the $x$ and $y$ coordinate system is a two-mode polarization-entangled state. It satisfies the eigenvalue relations

$$\hat{S}_0 |\psi; x, y\rangle = \hat{S}_2 |\psi; x, y\rangle = |\psi; x, y\rangle \tag{3.14}$$

with unit eigenvalues. Thus

$$\langle \hat{S}_0 \rangle = \langle \hat{n} \rangle = \langle \hat{S}_2 \rangle = 1 \quad \text{and} \quad V_0 = V_2 = 0. \tag{3.15}$$

The state considered is not an eigenstate of the remaining Stokes parameters, whose mean values are

$$\langle \hat{S}_1 \rangle = \langle \hat{S}_3 \rangle = 0. \tag{3.16}$$

However, the state does satisfy eigenvalue relations for the squares of these parameters, with

$$\hat{S}_1^2 |\psi; x, y\rangle = \hat{S}_3^2 |\psi; x, y\rangle = |\psi; x, y\rangle \tag{3.17}$$

and corresponding variances

$$V_1 = V_3 = 1. \tag{3.18}$$

It is readily verified that the three uncertainty relations in (2.8) are satisfied as equalities.

The state has unit total photon number and it is polarization squeezed in the $\hat{S}_2$ Stokes parameter in accordance with the criterion (3.12). The operators on both sides of (2.9), which define the Poincaré sphere, have eigenvalues equal to 3 for the entangled single-photon state.

The sphere is well-defined for this state, with a radius equal to $\sqrt{3}$. The sections of the Poincaré sphere shown in Fig. 1 apply to the entangled single-photon state if $n$ is set equal to 1 and the roles of the $S_1$ and $S_2$ axes are interchanged. The 45° rotation of the polarization leads to a 90° rotation on the Poincaré sphere. The corresponding angular momentum state is that of a spin $\frac{1}{2}$, as is discussed after (3.7).

*Two-mode quadrature-squeezed vacuum state*

The squeezed vacuum state of the two polarization modes is denoted

$$|\zeta;x,y\rangle = \hat{S}_{xy}(\zeta)|0\rangle_x|0\rangle_y \quad \text{where} \quad \zeta = se^{i\vartheta}. \tag{3.19}$$

Here

$$\hat{S}_{xy}(\zeta) = \exp\left(\zeta^*\hat{a}_x\hat{a}_y - \zeta\hat{a}_x^\dagger\hat{a}_y^\dagger\right) \tag{3.20}$$

is the usual two-mode squeeze operator [19], not to be confused with the Stokes parameters, with the properties

$$\begin{cases} \hat{S}_{xy}^\dagger(\zeta)\hat{a}_x\hat{S}_{xy}(\zeta) = \hat{a}_x\cosh s - \hat{a}_y^\dagger e^{i\vartheta}\sinh s \\ \hat{S}_{xy}^\dagger(\zeta)\hat{a}_y\hat{S}_{xy}(\zeta) = \hat{a}_y\cosh s - \hat{a}_x^\dagger e^{i\vartheta}\sinh s. \end{cases} \tag{3.21}$$

The mean photon numbers in the two modes are

$$\langle \hat{n}_x \rangle = \langle \hat{n}_y \rangle = \sinh^2 s. \tag{3.22}$$

The state is another example of an entangled state of the two polarization modes.

The two-mode quadrature-squeezed vacuum state satisfies the eigenvalue relation

$$\hat{S}_1|\zeta;x,y\rangle = 0, \tag{3.23}$$

which expresses the equality of the photon numbers in the two modes, and therefore

$$\langle \hat{S}_1 \rangle = 0 \quad \text{and} \quad V_1 = 0. \tag{3.24}$$

The mean values of the remaining Stokes parameters are

$$\langle \hat{S}_0 \rangle = \langle \hat{n} \rangle = 2\sinh^2 s \quad \text{and} \quad \langle \hat{S}_2 \rangle = \langle \hat{S}_3 \rangle = 0 \tag{3.25}$$

and their variances are

$$V_0 = V_2 = V_3 = \sinh^2 2s. \tag{3.26}$$

The mean values and variances of the Stokes parameters are all independent of the phase $\vartheta$ of the complex squeeze parameter $\zeta$. The two-mode quadrature-squeezed vacuum state is always polarization squeezed in $\hat{S}_1$ but not in $\hat{S}_2$ and $\hat{S}_3$. The expectation values of both sides of the Poincaré sphere relation (2.9) are equal to $2\sinh^2 2s$.

*Minimum-uncertainty amplitude-squeezed coherent states*

The state that has both polarization modes excited in identical but independent minimum-uncertainty amplitude-squeezed coherent states [20] is denoted

$$|\alpha,\zeta;x,y\rangle = \hat{D}_x(\alpha)\hat{S}_x(\zeta)\hat{D}_y(\alpha)\hat{S}_y(\zeta)|0\rangle_x|0\rangle_y \quad \text{where} \quad \zeta = se^{i\vartheta}. \tag{3.27}$$

The phase angle of the coherent complex amplitude $\alpha$ is equal to $\vartheta/2$ for amplitude squeezing. We assume, without loss of generality, that both these angles are zero. The squeeze parameter $\zeta$ then takes the real value $s$ and the squeeze operator is given by

$$\hat{S}_j(\zeta) = \exp\left\{\tfrac{1}{2}s\left[(\hat{a}_j)^2 - (\hat{a}_j^\dagger)^2\right]\right\}, \quad j = x, y. \tag{3.28}$$

The various required expectation values are evaluated by the standard methods [19]. Thus the mean photon numbers in the two modes are

$$\langle\hat{n}_x\rangle = \langle\hat{n}_y\rangle = \alpha^2 + \sinh^2 s. \tag{3.29}$$

The noise properties of the squeezed states are expressed in terms of the expectation values of the + and – quadrature operators, defined by

$$\hat{X}_j^+ = \hat{a}_j^\dagger + \hat{a}_j \quad \text{and} \quad \hat{X}_j^- = i\left(\hat{a}_j^\dagger - \hat{a}_j\right), \tag{3.30}$$

whose means and variances for the minimum-uncertainty amplitude-squeezed coherent states are

$$\langle\hat{X}_j^+\rangle = 2\alpha, \quad \langle\hat{X}_j^-\rangle = 0, \tag{3.31}$$

and

$$\langle(\delta\hat{X}_j^+)^2\rangle = e^{-2s}, \quad \langle(\delta\hat{X}_j^-)^2\rangle = e^{2s}, \tag{3.32}$$

where $j = x, y$ throughout. The + quadrature is squeezed and the – quadrature is anti-squeezed.

The expectation values of the Stokes parameters are

$$\langle \hat{S}_0 \rangle = 2\alpha^2 + 2\sinh^2 s \tag{3.33}$$

and

$$\langle \hat{S}_1 \rangle = \langle \hat{S}_3 \rangle = 0, \quad \langle \hat{S}_2 \rangle = 2\alpha^2. \tag{3.34}$$

The corresponding variances are

$$V_0 = V_1 = V_2 = 2\alpha^2 e^{-2s} + \sinh^2 2s \tag{3.35}$$

and

$$V_3 = 2\alpha^2 e^{2s}. \tag{3.36}$$

It is seen that the light may be separately squeezed or anti-squeezed in all of the Stokes parameters by appropriate choices of the values of $\alpha$ and $s$.

Much of the remainder of the present paper is concerned with bright amplitude-squeezed light, defined by $\alpha \gg \sinh s$. It is seen from (3.29), (3.35) and (3.36) that the light in this case is polarization squeezed in the $\hat{S}_1$ and $\hat{S}_2$ Stokes parameters and anti-squeezed in the $\hat{S}_3$ parameter, as is represented in Fig. 3. It is also squeezed in the $\hat{S}_0$ parameter, corresponding to photon-number squeezing.

## IV. BRIGHT AMPLITUDE-SQUEEZED LIGHT

We now discuss the production and measurement of polarization squeezing using a pair of bright amplitude-squeezed beams. Recently an effective method for producing such a pair of squeezed beams has been demonstrated experimentally [21,22]. In the following we will couch our discussion in terms of this technique for squeezing generation. However, any pair of amplitude-squeezed beams will exhibit similar properties. Like its predecessors, the proposed experiment, represented in Fig. 4, uses a fibre Sagnac interferometer followed by a polarizing beam splitter (PBS) [23] to produce two orthogonally-polarized amplitude-squeezed pulses, labeled $x$ and $y$. At the outputs, the two pulses are separated in time owing to the birefringence of the fibre but they can be brought into coincidence by an appropriate delay of the $y$ mode. The two pulses are then recombined at a second beam splitter. In a previous experiment [22], the polarization of the $x$ mode was rotated into the $y$ direction by a $\lambda/2$ plate, inserted after the first beam splitter; recombination by an ordinary beam splitter then produced two entangled output beams, both polarized in the $y$ direction. Because of thermal fluctuations, the optical phase of the interference had to be stabilized by a feed-back loop, controlled by equal DC signals of the detectors. By contrast, the version of this experiment considered here has no $\lambda/2$ plate and the pulses are brought together into one optical channel by a second PBS. This produces a single output beam with independent $x$ and $y$ contributions to the polarization. Whereas the independence of the quantum uncertainties provides the opportunity for polarization squeezing, the phase difference of the classical components must be regulated to obtain a single field with a defined polarization. Ideally, the light is guided into one output beam of the PBS but, in practice, imperfections in the PBS cause some fraction of each polarization to be lost to the other output. This loss effect can be used to implement the controller of a phase-locking loop. The polarization-squeezed beam so generated is referred to as the *primary beam*.

Following the analysis in [24], the mode operators for the primary beam are expressed as sums of identical real classical amplitudes $\alpha$ and quantum noise operators $\delta \hat{a}_j$,

$$\hat{a}_x = \alpha + \delta \hat{a}_x \quad \text{and} \quad \hat{a}_y = \alpha + \delta \hat{a}_y. \tag{4.1}$$

The expectation values of the noise operators are assumed to be much smaller than the coherent amplitude $\alpha$. Then, to first order in the $\delta \hat{a}_j$, the Stokes operators from (2.1) to (2.4) are

$$\hat{S}_0 = 2\alpha^2 + \alpha(\delta \hat{X}_x^+ + \delta \hat{X}_y^+) \quad \langle \hat{S}_0 \rangle = 2\alpha^2 \tag{4.2}$$

$$\hat{S}_1 = \alpha(\delta \hat{X}_x^+ - \delta \hat{X}_y^+) \quad \langle \hat{S}_1 \rangle = 0 \tag{4.3}$$

$$\hat{S}_2 = 2\alpha^2 + \alpha(\delta \hat{X}_x^+ + \delta \hat{X}_y^+) \quad \langle \hat{S}_2 \rangle = 2\alpha^2 \tag{4.4}$$

$$\hat{S}_3 = -\alpha(\delta \hat{X}_x^- - \delta \hat{X}_y^-) \quad \langle \hat{S}_3 \rangle = 0, \tag{4.5}$$

where the quadrature operators are defined in (3.30). Their mean values agree with those of the minimum-uncertainty squeezed states in (3.33) and (3.34) when $\alpha \gg \sinh s$; they are also the same as those for identical coherent-state excitations in the two polarization modes.

The variances of the Stokes parameters are

$$V_0 = V_1 = V_2 = \alpha^2 \{\langle (\delta \hat{X}_x^+)^2 \rangle + \langle (\delta \hat{X}_y^+)^2 \rangle\} \tag{4.6}$$

$$V_3 = \alpha^2 \{\langle (\delta \hat{X}_x^-)^2 \rangle + \langle (\delta \hat{X}_y^-)^2 \rangle\}. \tag{4.7}$$

These expressions apply for arbitrary values of the quadrature-operator variances. In the special case of minimum-uncertainty amplitude-squeezed coherent states with $\alpha \gg \sinh s$, they agree with the variances obtained from (3.32), (3.35) and (3.36). More generally, for amplitude-squeezed states that do not satisfy the minimum-uncertainty condition, polarization squeezing of the primary beam may still occur for the Stokes parameters in (4.6).

The four Stokes parameters in classical optics are measured by well-known techniques that involve transmission of the light beam of interest through appropriate combinations of quarter-wave plate with polarization rotators [2,4,23]. The parameters are then obtained by measurements of the intensities of orthogonally-polarized components of the output light. These measurements are usually made successively, first on one polarization and then on the other. The mean values of the quantum Stokes parameters are similarly measurable by direct detection after appropriate processing of the incident light. However, more care is needed in the measurement of the quantum noise properties of the Stokes parameters. It is clear from (2.1) and (2.2) that the statistical properties of the observed Stokes parameters $\hat{S}_0$ and $\hat{S}_1$, including their means and variances, can be obtained from the sum and difference of the directly-detected photon numbers in the $x$ and $y$ components of the primary beam. The experimental setup for their detection is shown in Fig. 5. These Stokes parameters essentially describe properties of the individual polarization components and their photon-number squeezing.

Consider, however, the quantum properties of the $\hat{S}_2$ parameter. The classical measurement of this parameter is made by subtraction of the directly-detected intensities of the beam after its passage through polarizers successively oriented at $45°$ and $-45°$ to the $x$ axis. With use of the Jones matrices [23], a rotation of the polarizations through $45°$ with respect to the $x$ axis converts the mode operators to new primed axes in accordance with

$$\begin{bmatrix} \hat{a}_{x'} \\ \hat{a}_{y'} \end{bmatrix} = \begin{bmatrix} 2^{-1/2} & 2^{-1/2} \\ -2^{-1/2} & 2^{-1/2} \end{bmatrix} \begin{bmatrix} \hat{a}_x \\ \hat{a}_y \end{bmatrix}. \tag{4.8}$$

However, in contrast to the classical procedure, where separate measurements are made on the two polarization components, the polarization-rotated beam is here sent into the $\hat{a}$ arm of the PBS represented in Fig. 6. The transmission and reflection axes of the PBS are oriented parallel to the primed axes and its input–output relations are

$$\begin{bmatrix} \hat{c}_{x'} \\ \hat{c}_{y'} \\ \hat{d}_{x'} \\ \hat{d}_{y'} \end{bmatrix} = \begin{bmatrix} & & 1 & \\ & 1 & & \\ 1 & & & \\ & & & 1 \end{bmatrix} \begin{bmatrix} \hat{a}_{x'} \\ \hat{a}_{y'} \\ \hat{b}_{x'} \\ \hat{b}_{y'} \end{bmatrix}. \tag{4.9}$$

With no input to the $\hat{b}$ arm of the beam splitter, so that both $\hat{b}$ polarization modes are in their vacuum states, the input–output relations are conveniently inverted to give

$$\begin{bmatrix} \hat{a}_x \\ \hat{a}_y \\ \hat{b}_{x'} \\ \hat{b}_{y'} \end{bmatrix} = \begin{bmatrix} -2^{-1/2} & 2^{-1/2} & & \\ 2^{-1/2} & 2^{-1/2} & & \\ & & 1 & \\ & & & 1 \end{bmatrix} \begin{bmatrix} \hat{c}_{x'} \\ \hat{c}_{y'} \\ \hat{d}_{x'} \\ \hat{d}_{y'} \end{bmatrix}. \tag{4.10}$$

It is readily shown that

$$\hat{S}_2 = \hat{a}_x^\dagger \hat{a}_y + \hat{a}_y^\dagger \hat{a}_x = \hat{d}_{x'}^\dagger \hat{d}_{x'} - \hat{c}_{y'}^\dagger \hat{c}_{y'}, \tag{4.11}$$

in accordance with the definition in (2.3). The $\hat{S}_2$ Stokes parameter of the primary beam is thus obtained by taking the difference between direct-detection measurements of the respective $x'$ and $y'$ polarization components in the two output arms of the PBS. It is emphasized that, in general, simultaneous measurements are made on the individual pulses that make up the primary beam. These measurements provide experimental values for the mean and the variance calculated in (4.4) and (4.6) respectively. However, for the assumed bright beams in both the $\hat{a}$ inputs, the $\hat{c}_{y'}$ output is dark and it can be neglected. The output $\hat{d}_{x'}$ is bright with the intensity variance equal to $V_2$. Figure 7 shows the corresponding experimental setup.

A measurement of the $\hat{S}_3$ parameter is made by a variant of the above procedure in which a quarter-wave plate is inserted into the primary beam before polarization rotation. With use of the appropriate Jones matrix [23], the inverted input–output relations are now

$$\begin{bmatrix} \hat{a}_x \\ \hat{a}_y \\ \hat{b}_{x'} \\ \hat{b}_{y'} \end{bmatrix} = \begin{bmatrix} -2^{-1/2} & 2^{-1/2} & & \\ 2^{-1/2}i & 2^{-1/2}i & & \\ & & 1 & \\ & & & 1 \end{bmatrix} \begin{bmatrix} \hat{c}_{x'} \\ \hat{c}_{y'} \\ \hat{d}_{x'} \\ \hat{d}_{y'} \end{bmatrix} \tag{4.12}$$

and it is easily shown that

$$\hat{S}_3 = i\left(\hat{a}_y^\dagger \hat{a}_x - \hat{a}_x^\dagger \hat{a}_y\right) = \hat{d}_{x'}^\dagger \hat{d}_{x'} - \hat{c}_{y'}^\dagger \hat{c}_{y'}, \tag{4.13}$$

in accordance with the definition in (2.4). The final Stokes parameter is thus again measured by taking the difference of two direct-detection measurements on the PBS outputs. However, in contrast to the measurement of $\hat{S}_2$ for bright $\hat{a}$ inputs, both output beams $\hat{c}_{y'}$ and $\hat{d}_{x'}$ are now bright and actually contribute to the variance of the Stokes parameter $\hat{S}_3$. The detection scheme is depicted in Fig. 8. Note that the vacuum operators $\hat{b}_{x'}$ and $\hat{b}_{y'}$ affect only the unobserved $\hat{c}_{x'}$ and $\hat{d}_{y'}$ output modes for the measurements of both $\hat{S}_2$ and $\hat{S}_3$.

In the special case where the two components of the primary beam are excited in minimum-uncertainty amplitude-squeezed coherent states, the joint squeeze operator from (3.28) is

$$\hat{S}_x(\zeta)\hat{S}_y(\zeta) = \exp\left\{\frac{1}{2}s\left[(\hat{a}_x)^2 + (\hat{a}_y)^2 - (\hat{a}_x^\dagger)^2 - (\hat{a}_y^\dagger)^2\right]\right\}$$
$$= \exp\left\{s\left(\hat{c}_{y'}\hat{d}_{x'} - \hat{c}_{y'}^\dagger\hat{d}_{x'}^\dagger\right)\right\} \quad (4.14)$$

This has the same form as the operator in (3.20) and the $\hat{c}_{y'}$ and $\hat{d}_{x'}$ outputs from the PBS are thus excited in an entangled two-mode squeezed coherent state.

The above analysis shows that the variances of the Stokes parameters are closely related to the quadrature variances. Thus measurements of the Stokes parameters essentially determine the quadrature variances by appropriate manipulations of the two polarization components of the primary beam. All three of the Stokes measurements involve only direct detection and there is no need for the local oscillator normally used in phase-sensitive observations of the quadrature squeezing. In the experiments proposed here, the two polarization components of the primary beam essentially replace the squeezed signal and coherent local oscillator of the conventional squeezing measurement.

# V. POLARIZATION EPR STATES

We now consider the analogue of Einstein, Podolsky, Rosen (EPR) entanglement [25] for the quantum Stokes parameters. Suppose we combine two independent polarization-squeezed primary beams, of the type discussed in the previous section, on an ordinary beam splitter, analogous to the scheme presented in [24]. Suppose also that we impose a phase shift $\pi/2$ on one of them before letting them interfere on a beam splitter. The relations between input and output mode operators of the beam splitter have then the forms shown in Fig. 9. The phases of the combinations are thus arranged so that squeezed and anti-squeezed quadratures of the various beams are superimposed. This is a direct generalization of the method used to generate standard EPR entanglement [26]. The mode operators of the output beams, $\hat{c}_x$, $\hat{c}_y$ and $\hat{d}_x$, $\hat{d}_y$, are given by

$$\hat{c}_x = \frac{1}{\sqrt{2}}\left[(1+i)\alpha + \delta\hat{a}_x + i\delta\hat{b}_x\right], \quad \hat{c}_y = \frac{1}{\sqrt{2}}\left[(1+i)\alpha + \delta\hat{a}_y + i\delta\hat{b}_y\right],$$
$$\hat{d}_x = \frac{1}{\sqrt{2}}\left[(1-i)\alpha + \delta\hat{a}_x - i\delta\hat{b}_x\right], \quad \hat{d}_y = \frac{1}{\sqrt{2}}\left[(1-i)\alpha + \delta\hat{a}_y - i\delta\hat{b}_y\right], \quad (5.1)$$

where the bright beams in the $\hat{a}$ and $\hat{b}$ inputs have identical real classical amplitudes $\alpha$ plus quantum noise operators, similar to (4.1). Such beams have been considered before as a continuous variable teleportation resource [27] and for creating Bell-type correlations for continuous variables [28]. In both cases, quadrature amplitude measurements employing local oscillators were employed. Here we will show that the Stokes parameters of the two beams, directly measureable as previously described, satisfy the standard EPR condition for entanglement.

EPR entanglement is defined to occur when measurements carried out on one subsystem can be used to infer the values of non-commuting observables of another, spatially-separated subsystem to sufficient precision that an "apparent" violation of the uncertainty principle occurs [29]. The precision with which we can infer the value of an observable $\hat{Z}_D$ of subsystem $D$ from the measurement of $\hat{Z}_C$ on subsystem $C$ is given by the conditional variance

$$V_{\text{cond}}(Z_D|Z_C) = \langle (\delta\hat{Z}_D)^2 \rangle - \frac{|\langle \delta\hat{Z}_D \delta\hat{Z}_C \rangle|^2}{\langle (\delta\hat{Z}_C)^2 \rangle}. \tag{5.2}$$

Then EPR entanglement of the Stokes parameters will be realized, for example, if

$$V_{\text{cond}}(S_{3D}|S_{3C}) V_{\text{cond}}(S_{1D}|S_{1C}) < |\langle \hat{S}_{2C} \rangle|^2. \tag{5.3}$$

Here the output $\hat{c}$ in Fig. 9 is assigned to the subsystem $C$ and the output $\hat{d}$ to subsystem $D$. Now, after performing the linearization, we have that

$$\begin{aligned}
\delta\hat{S}_{1C} &= \hat{c}_x^\dagger \hat{c}_x - \hat{c}_y^\dagger \hat{c}_y \\
&= \tfrac{1}{2}\alpha\left(\delta\hat{X}_{ax}^+ + \delta\hat{X}_{ax}^- - \delta\hat{X}_{ay}^+ - \delta\hat{X}_{ay}^- + \delta\hat{X}_{bx}^+ - \delta\hat{X}_{bx}^- - \delta\hat{X}_{by}^+ + \delta\hat{X}_{by}^-\right)
\end{aligned} \tag{5.4}$$

where indices $ax, ay, bx, by$ are related to the corresponding polarization modes $\hat{a}_j$ and $\hat{b}_j$, $j = x, y$ and $\delta\hat{S}_{1D}$ is given by the same expression but with the signs of all contributions $\delta\hat{X}^-$ reversed. Similarly

$$\begin{aligned}
\delta\hat{S}_{3C} &= i\left(\hat{c}_y^\dagger \hat{c}_x - \hat{c}_x^\dagger \hat{c}_y\right) \\
&= \tfrac{1}{2}\alpha\left(\delta\hat{X}_{ax}^+ - \delta\hat{X}_{ax}^- - \delta\hat{X}_{ay}^+ + \delta\hat{X}_{ay}^- - \delta\hat{X}_{bx}^+ - \delta\hat{X}_{bx}^- + \delta\hat{X}_{by}^+ + \delta\hat{X}_{by}^-\right)
\end{aligned} \tag{5.5}$$

and $\delta\hat{S}_{3D}$ is given by the same expression but with the signs of all contributions $\delta\hat{X}^+$ reversed. It follows from these expressions that

$$\begin{aligned}
\langle (\delta\hat{S}_{1C})^2 \rangle &= \langle (\delta\hat{S}_{3C})^2 \rangle = \langle (\delta\hat{S}_{1D})^2 \rangle = \langle (\delta\hat{S}_{3D})^2 \rangle \\
&= \tfrac{1}{4}\alpha^2 \left\{V_{ax}^+ + V_{ax}^- + V_{ay}^+ + V_{ay}^- + V_{bx}^+ + V_{bx}^- + V_{by}^+ + V_{by}^-\right\}
\end{aligned} \tag{5.6}$$

and

$$\langle \delta\hat{S}_{1D}\delta\hat{S}_{1C}\rangle = -\langle \delta\hat{S}_{3D}\delta\hat{S}_{3C}\rangle$$
$$= \tfrac{1}{4}\alpha^2\{V_{ax}^+ - V_{ax}^- + V_{ay}^+ - V_{ay}^- + V_{bx}^+ - V_{bx}^- + V_{by}^+ - V_{by}^-\}, \quad (5.7)$$

where, for example, $V_{ax}^\pm = \langle(\delta X_{ax}^\pm)^2\rangle$. Finally

$$|\langle\hat{S}_{2C}\rangle|^2 = 4\alpha^4. \quad (5.8)$$

The conditional variances are thus

$$V_{\text{cond}}(S_{1D}|S_{1C}) = V_{\text{cond}}(S_{3D}|S_{3C})$$
$$= \tfrac{1}{4}\alpha^2\left(V_{ax}^+ + V_{ax}^- + V_{ay}^+ + V_{ay}^- + V_{bx}^+ + V_{bx}^- + V_{by}^+ + V_{by}^-\right) \quad (5.9)$$
$$- \frac{\alpha^2\left(V_{ax}^+ - V_{ax}^- + V_{ay}^+ - V_{ay}^- + V_{bx}^+ - V_{bx}^- + V_{by}^+ - V_{by}^-\right)^2}{4\left(V_{ax}^+ + V_{ax}^- + V_{ay}^+ + V_{ay}^- + V_{bx}^+ + V_{bx}^- + V_{by}^+ + V_{by}^-\right)}.$$

These expressions can be used to assess the EPR entanglement condition in (5.3).

If we assume that the modes making up the original polarization-squeezed beams all have equal quadrature squeezing, that is

$$V_{ax}^+ = V_{ay}^+ = V_{bx}^+ = V_{by}^+ = V^+ \text{ and } V_{ax}^- = V_{ay}^- = V_{bx}^- = V_{by}^- = V^-, \quad (5.10)$$

then we obtain from (5.9)

$$V_{\text{cond}}(S_{1D}|S_{1C}) = V_{\text{cond}}(S_{3D}|S_{3C}) = 4\alpha^2 \frac{V^+V^-}{V^+ + V^-}. \quad (5.11)$$

Compare now (5.3), (5.8), and (5.11). For minimum-uncertainty quadrature squeezed modes, where $V^+V^- = 1$ as in (3.32), any level of squeezing will lead to the Stokes EPR condition (5.3) being satisfied. Non-minimum uncertainty states must fulfil more stringent squeezing conditions [30], although there remain ranges of values of $V^-$ (or $V^+$) for which (5.3) is satisfied when $V^+ < 1$ (or $V^- < 1$). For example, when $V^+ < 1/2$, which corresponds to 3 dB of squeezing, Stokes EPR entanglement occurs for all values of $V^-$ in the range $1/V^+ \leq V^- < \infty$. We propose that these polarization-entangled EPR states can be usefully employed to implement continuous-variable quantum information protocols in the absence of a local oscillator.

Another way to analyze our polarization entanglement states is to use the continuous variable Peres-Horodecki criterion for separability. This criterion verifies whether two subsystems D and C are entangled [31]. For two pairs of conjugate variables $\hat{Z}_D, \hat{W}_D$ and $\hat{Z}_C, \hat{W}_C$ of these subsystems the criterion can be written in the form:

$$V_{\pm}(Z_D, Z_C) + V_{\mp}(W_D, W_C) < 2 \tag{5.12}$$

where the relevant variances are defined as:

$$V_{\pm}(Z_D, Z_C) = \frac{V(\hat{Z}_D \pm \hat{Z}_C)}{V(\hat{Z}_D^{coh} + \hat{Z}_C^{coh})}, \qquad V_{\mp}(W_D, W_C) = \frac{V(\hat{W}_D \mp \hat{W}_C)}{V(\hat{W}_D^{coh} + \hat{W}_C^{coh})} \tag{5.13}$$

and the values labeled "coh" correspond to respective coherent states. In the spirit of this non-separability criterion [31], we define the following entanglement boundary for the Stokes operators $\hat{S}_1$ and $\hat{S}_3$:

$$V_{\pm}(S_{1D}, S_{1C}) = \frac{V(\hat{S}_{1D} \pm \hat{S}_{1C})}{V(\hat{S}_{1D}^{coh} + \hat{S}_{1C}^{coh})} < 1, \qquad V_{\mp}(S_{3D}, S_{3C}) = \frac{V(\hat{S}_{3D} \mp \hat{S}_{3C})}{V(\hat{S}_{3D}^{coh} + \hat{S}_{3C}^{coh})} < 1; \tag{5.14}$$

or for $\hat{S}_2$ and $\hat{S}_3$, analogously:

$$V_{\pm}(S_{2D}, S_{2C}) = \frac{V(\hat{S}_{2D} \pm \hat{S}_{2C})}{V(\hat{S}_{2D}^{coh} + \hat{S}_{2C}^{coh})} < 1, \qquad V_{\mp}(S_{3D}, S_{3C}) = \frac{V(\hat{S}_{3D} \mp \hat{S}_{3C})}{V(\hat{S}_{3D}^{coh} + \hat{S}_{3C}^{coh})} < 1. \tag{5.15}$$

Here a more stringent condition is introduced as compared to the one used in [31]. It requires the variances of *both* conjugate variables $\hat{S}_1$ and $\hat{S}_3$ ( or $\hat{S}_2$ and $\hat{S}_3$, etc) to drop below the limit imposed by the continuous variable version of the Peres-Horodecki criterion [31]. We refer to entanglement satisfying (5.14) as *squeezed-state entanglement* [30]. Such a non-separability condition (5.12) in its modified form (5.14) is important for the application of entanglement in quantum communication protocols. Both conjugate variables have to exhibit a quantum correlation to guarantee secure quantum key distribution. A quantum correlation of both conjugate variables is also preferable for the reconstruction of an unknown state in quantum teleportation. The squeezed-state entanglement has the nature of a two-mode squeezed state, hence its name. For large squeezing levels it satisfies also the EPR criterion (5.3). Note, however, that in this case again *both* variances building the EPR product (5.3) fall below the respective limit of $|\langle S_j \rangle|$ ( $j = 1$ for $\hat{S}_2$ and $\hat{S}_3$, $j = 2$ for $\hat{S}_1$ and $\hat{S}_3$, etc) in contrast to the original definition [29] and (5.3).

The values of the variances for coherent bright beams can be calculated using expressions (3.9), (3.10) and (4.2)-(4.7) which delivers $V(\hat{S}_{jD}^{coh} + \hat{S}_{jC}^{coh}) = 4\alpha^2$, $j = 1, 2, 3$. If we again assume that the modes making up the original polarization-squeezed beams all have the equal squeezing (5.10), then for the bright beam example described above we get the squeezing variances (5.13), (5.14) of

$$V_+(S_{1D}, S_{1C}) = \frac{V(\hat{S}_{1D} + \hat{S}_{1C})}{V(\hat{S}_{1D}^{coh} + \hat{S}_{1C}^{coh})} = V^+, \qquad V_-(S_{3D}, S_{3C}) = \frac{V(\hat{S}_{3D} - \hat{S}_{3C})}{V(\hat{S}_{3D}^{coh} + \hat{S}_{3C}^{coh})} = V^+. \qquad (5.16)$$

The criterion of squeezed-state entanglement is thus always satisfied for the input amplitude-squeezed beams with $V_{ax}^+ = V_{ay}^+ = V_{bx}^+ = V_{by}^+ = V^+ < 1$.

Note that the Peres-Horodecki criterion (5.12):

$$V_+(S_{1D}, S_{1C}) + V_-(S_{3D}, S_{3C}) = V_a^+ + V_b^+ < 2 \qquad (5.17)$$

is satisfied also when only one of the input fields $V_{ay}^+ = V_{ay}^+ = V_a^+$; $V_{bx}^+ = V_{by}^+ = V_b^+$ exhibits amplitude quadrature squeezing ($V_a^+ < 1$ or $V_b^+ < 1$), the other one being coherent. Hence a non-separable two-mode field is generated in the interference of one single polarization-squeezed beam with a coherent (or vacuum) one on a beam splitter.

The experimental setup for the generation of bright beams quantum-correlated in polarization is represented in Fig. 10. Quantum correlations between the uncertainties of the Stokes operators already emerge in the interference of a polarization-squeezed beam with a vacuum or coherent field in the other input of a beam splitter, as is shown in Sec. IV and as follows from (5.12), (5.17). However, taking into account the realistic squeezing levels of the input fields, achievable in an experiment, the interference of two polarization-squeezed beams is needed to produce the continuous variable polarization EPR entanglement.

## VI. CONCLUSIONS

Both the classical and quantum Stokes parameters represent useful tools for the description of the polarization of a light beam and also, more generally, of the phase properties of two-mode fields. They include explicitly the phase difference between the modes and they can be reliably measured in experiments. These features have triggered the use of Stokes operators for the construction of a formalism for the quantum description of relative phase [15]. The novel approach to quantum information processing that exploits quantum continuous variables has also stimulated interest in the nonclassical polarization states. The formalism of the quantum Stokes operators was recently used to describe the mapping of the polarization state of a light beam on to the spin variables of atoms in excited states [32,33]; the correspondence between the algebras of the Stokes operators and the spin operators enables an efficient transfer of quantum information from a freely-propagating optical carrier to a matter system. These developments pave the way towards the quantum teleportation of atomic states and towards the storage and read-out of quantum information. In the present paper, we have applied the concepts of quantum Stokes operators and nonclassical polarization states to schemes for quantum-optical communication with bright squeezed light beams.

The basic properties of the quantum Stokes operators are reviewed in Sec. II and they are illustrated in Sec. III by applications to a range of simple quantum-mechanical polarization states. Polarization squeezing, defined as the occurrence of variances in one or more of the Stokes parameters smaller than the coherent-state value, is found in photon-number states, entangled single-photon states, the two-mode quadrature-squeezed vacuum state and the minimum-uncertainty amplitude-squeezed coherent states. For many practical applications, it is preferable to use the bright amplitude-squeezed light that is available experimentally and this is considered in Sec. IV. It is shown in particular how all of the Stokes parameters can be measured by the use of linear optical elements and direct detection schemes that are sensitive to the quantum correlations in the two polarization components of the light. These measurement schemes are developments of the well-known methods for determination of the classical Stokes parameters to preserve the quantum noise properties.

The continuous-variable polarization EPR entanglement considered in Sec. V implies correlations between the quantum uncertainties of a pair of Stokes operators as conjugate variables. The entanglement can be generated by linear interference of two polarization-squeezed beams on a beam splitter and the relevant conjugate variables are measured as before by direct detection schemes. We propose to apply bright polarization-entangled beams to continuous-variable quantum cryptography [34, 35] where it allows one to dispense with the experimentally costly local oscillator techniques. Implementation of the protocols [34, 35] using continuous-variable polarization entanglement combines the advantages of intense easy-to-handle sources of EPR-entangled light and efficient direct detection, thus opening the way to secure quantum communication with bright light. In general, we believe that nonclassical polarization states can be used with advantage in quantum information protocols that involve measurements of both conjugate continuous variables and in quantum state transfer from light fields to matter systems.

## ACKNOWLEDGEMENTS

The work was supported by the Deutsche Forschungsgemeinschaft and by the EU grant under QIPC, project IST-1999-13071 (QUICOV). RL acknowledges the financial support of the Alexander von Humboldt Foundation. TCR acknowledges useful discussions with Prem Kumar and the support of the Australian Research Council. RL and TCR greatly appreciate the warm hospitality of Professor Gerd Leuchs and his group in Erlangen.

# FIGURE CAPTIONS

1.  Sections of the Poincaré sphere in the 31 and 23 planes for the $x$ polarized number state. The heavy points and the circle respectively show the locus of the tip of the Stokes vector in these planes.

2.  Sections of the Poincaré sphere in the 31 and 23 planes for the $x$ polarized coherent state. The shaded discs show the projections of the uncertainty sphere of the Stokes vector in these planes.

3.  Representations of quantum polarization states of bright coherent and bright amplitude-squeezed light on the Poincaré sphere. The latter shows polarization squeezing in the parameters $\hat{S}_0$, $\hat{S}_1$, and $\hat{S}_2$, with anti-squeezing in $\hat{S}_3$.

4.  Experimental setup for the generation of bright polarization-squeezed light. VA: variable attenuator, $\lambda/2$: half-wave plate, 90/10: beam splitter with 90% reflectivity, PBS: polarizing beam splitter. The two orthogonal polarizations from the Sagnac interferometer are labeled $x$ and $y$.

5.  Scheme for measurement of the Stokes parameters $\hat{S}_0$ and $\hat{S}_1$.

6.  Polarizing beam splitter showing the notation for input and output modes.

7.  Scheme for measurement of the Stokes parameter $\hat{S}_2$ for bright beams.

8.  Scheme for measurement of the Stokes parameter $\hat{S}_3$.

9.  Interference of two bright polarization-squeezed beams on a beam splitter.

10. Experiment for the generation of continuous variable EPR polarization-entangled states.

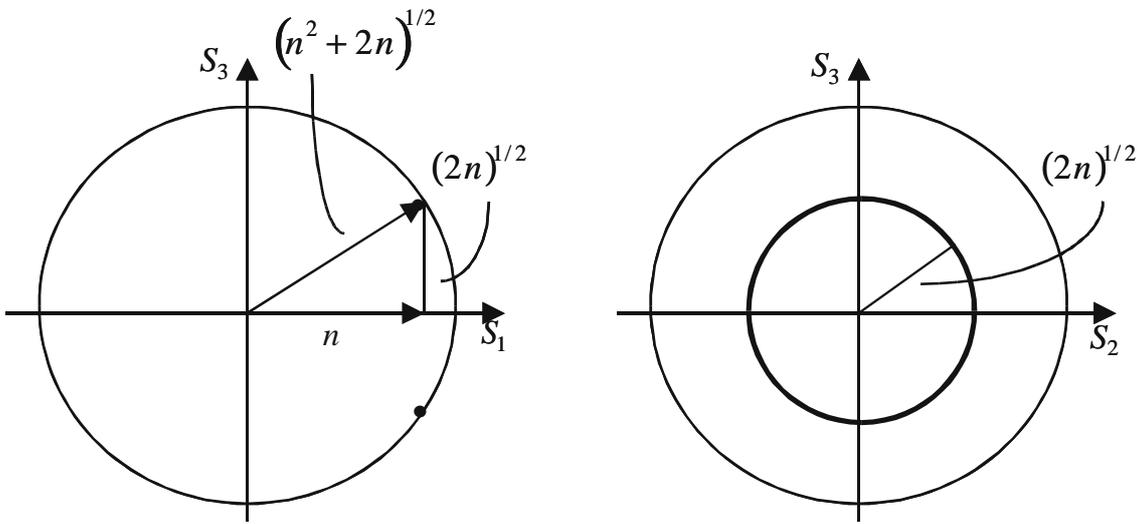

Figure 1

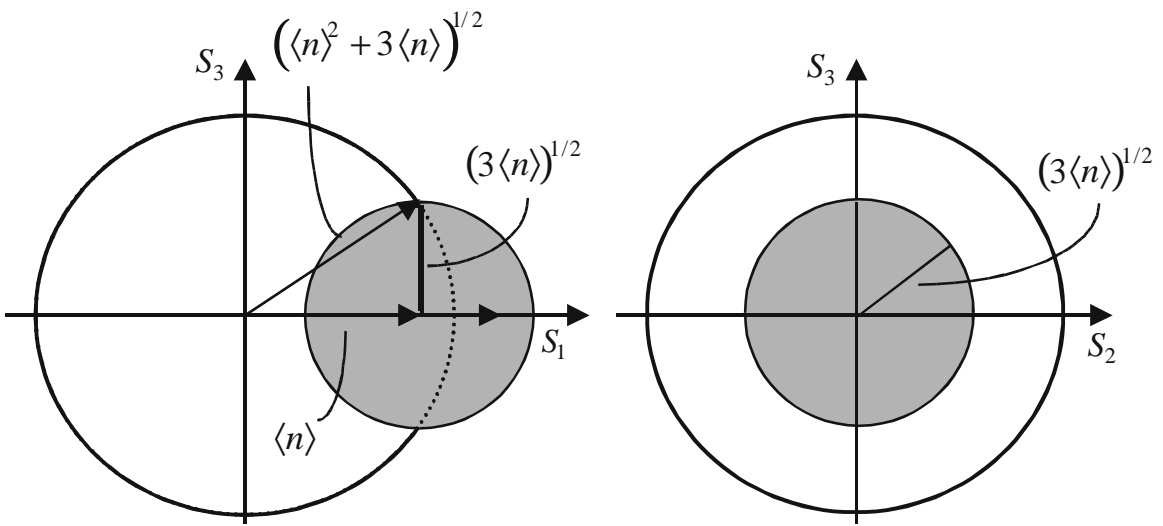

Figure 2

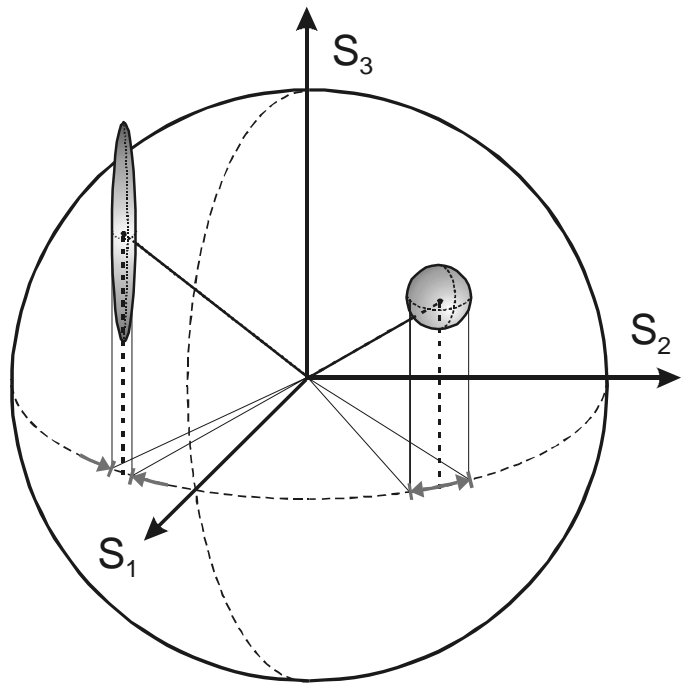

Figure 3

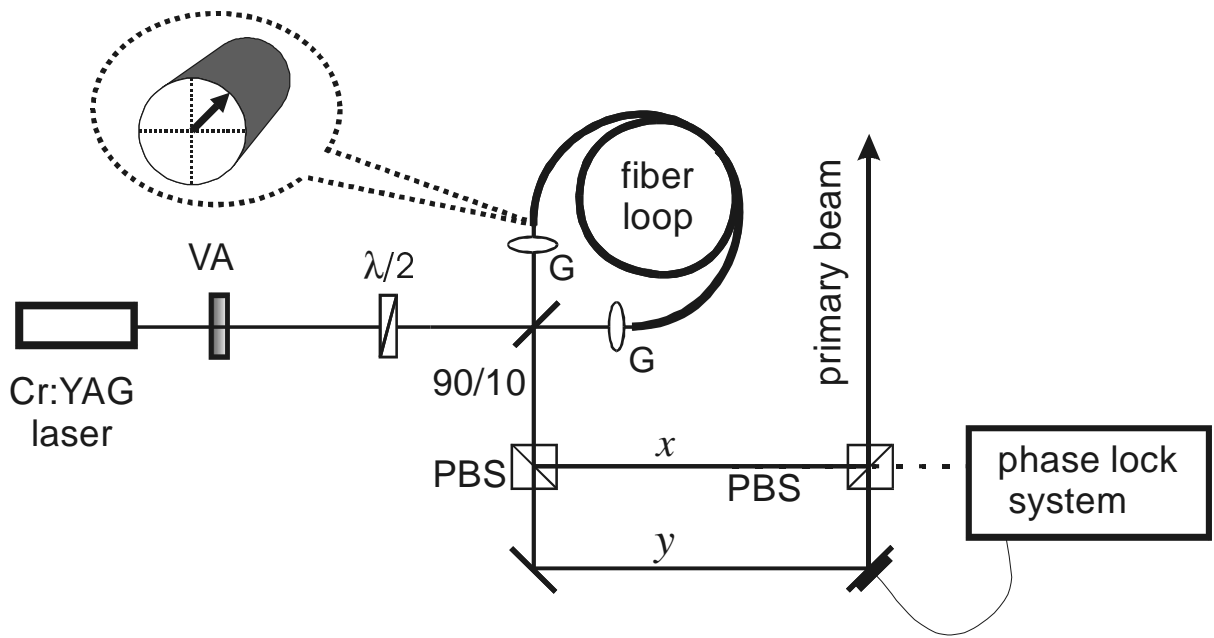

Figure 4

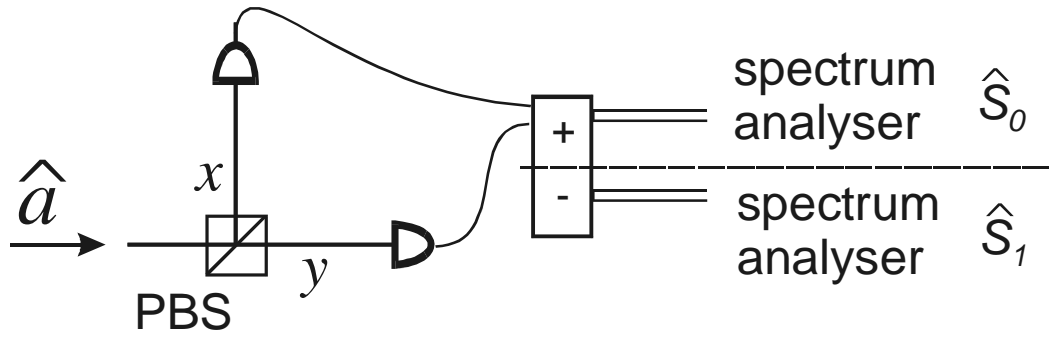

Figure 5

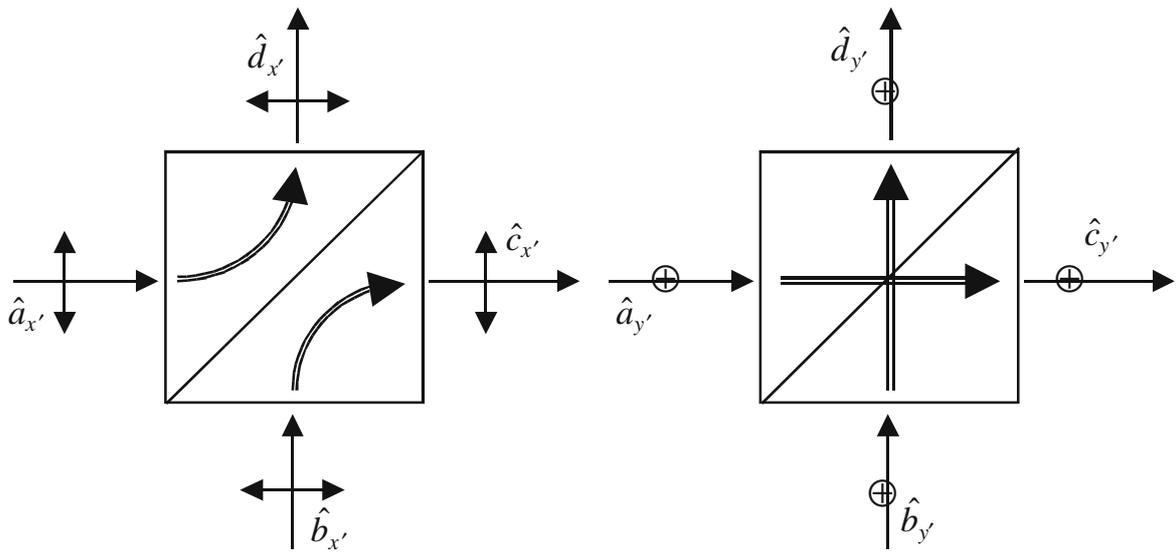

Figure 6

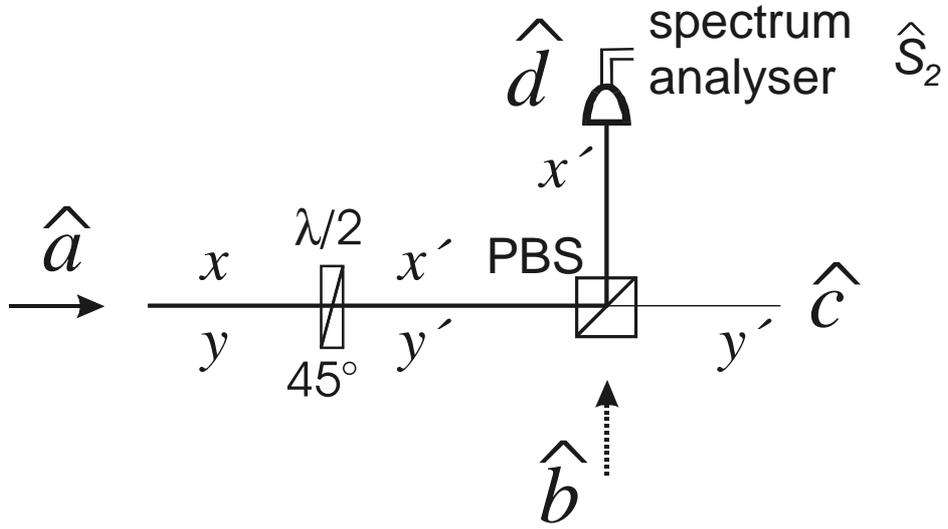

Figure 7

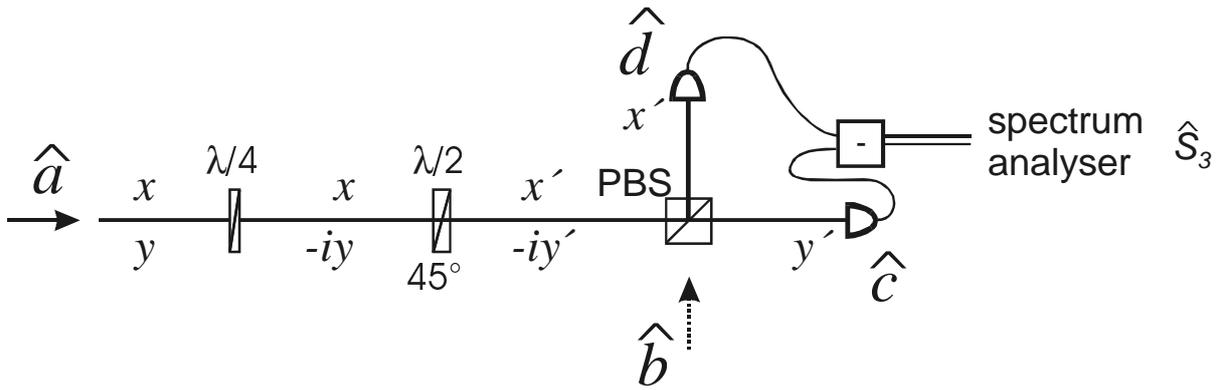

Figure 8

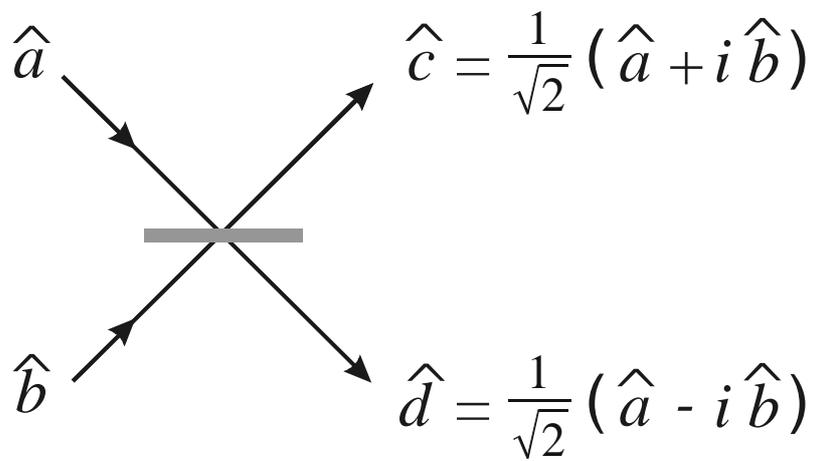

Figure 9

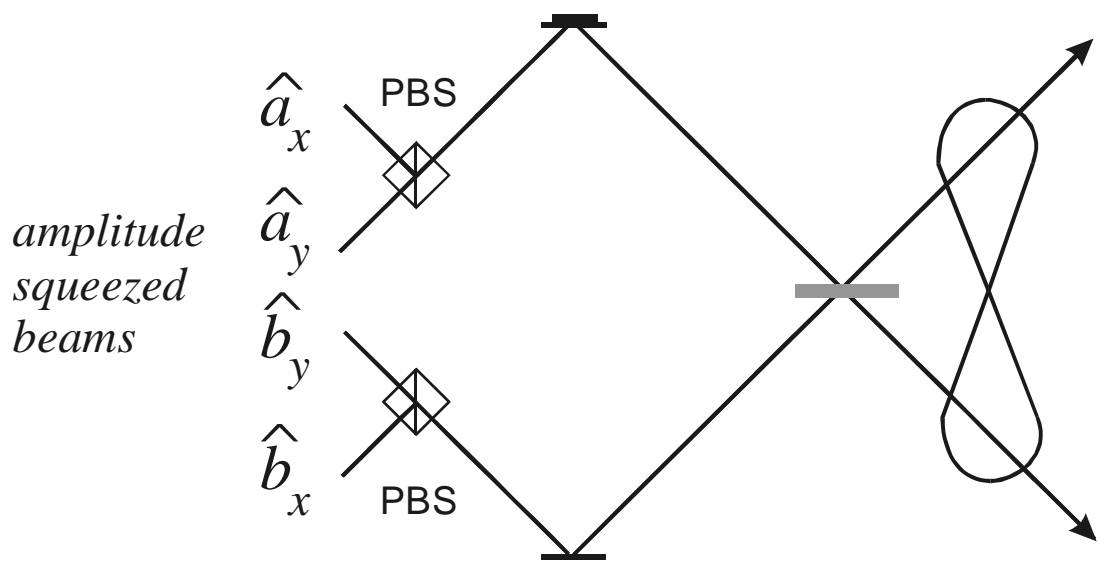

Figure 10